\documentclass[12pt,a4paper]{article}

\renewcommand{\theequation}{\arabic{equation}}
\newcommand{\EQ}{\begin{equation}}
\newcommand{\EN}{\end{equation}}

\newcommand{\bear}{\begin{eqnarray}}
\newcommand{\ear}{\end{eqnarray}}
\newcommand{\bt} { \begin{tabular} }
\newcommand{\et}{ \end{tabular} }
\newcommand{\bc} { \begin{center} }
\newcommand{\ec}{ \end{center} }

\newcommand{\btb} { \begin{table} }
\newcommand{\etb}{ \end{table} }

\begin{document}

\topmargin 0pt
\oddsidemargin 5mm
\newcommand{\NP}[1]{Nucl.\ Phys.\ {\bf #1}}
\newcommand{\PL}[1]{Phys.\ Lett.\ {\bf #1}}
\newcommand{\NC}[1]{Nuovo Cimento {\bf #1}}
\newcommand{\CMP}[1]{Comm.\ Math.\ Phys.\ {\bf #1}}
\newcommand{\PR}[1]{Phys.\ Rev.\ {\bf #1}}
\newcommand{\PRL}[1]{Phys.\ Rev.\ Lett.\ {\bf #1}}
\newcommand{\MPL}[1]{Mod.\ Phys.\ Lett.\ {\bf #1}}
\newcommand{\JETP}[1]{Sov.\ Phys.\ JETP {\bf #1}}
\newcommand{\TMP}[1]{Teor.\ Mat.\ Fiz.\ {\bf #1}}

\renewcommand{\thefootnote}{\fnsymbol{footnote}}

\newpage
\setcounter{page}{0}
\begin{titlepage}
\begin{flushright}
UFSCARF-TH-04-15
\end{flushright}
\vspace{0.5cm}
\begin{center}
{\large New $R$-matrices from Representations of Braid-Monoid\\  Algebras based on Superalgebras}\\
\vspace{1cm}
{\large W. Galleas and M.J. Martins } \\
\vspace{1cm}
{\em Universidade Federal de S\~ao Carlos\\
Departamento de F\'{\i}sica \\
C.P. 676, 13565-905~~S\~ao Carlos(SP), Brasil}\\
\end{center}
\vspace{0.5cm}

\begin{abstract}
In this paper we discuss representations of the Birman-Wenzl-Murakami 
algebra as well as of its  dilute extension  containing several free parameters. 
These representations are based on superalgebras and their baxterizations permit us to derive novel
trigonometric solutions of the graded Yang-Baxter equation. In this way we obtain
the multiparametric $R$-matrices associated to the 
$U_q[sl(r|2m)^{(2)}]$, $U_q[osp(r|2m)^{(1)}]$ and 
$U_q[osp(r=2n|2m)^{(2)}]$ quantum symmetries. 
Two other families
of multiparametric $R$-matrices not predicted before 
within the context of quantum superalgebras are also presented. The latter systems
are indeed 
non-trivial generalizations 
of the 
$U_q[D^{(2)}_{n+1}]$  vertex model 
when both distinct edge variables statistics 
and extra free-parameters
are admissible.

\end{abstract}

\vspace{.15cm}
\centerline{PACS numbers:  05.50+q, 02.30.IK}
\vspace{.1cm}
\centerline{Keywords:  Integrable lattice models, Yang-Baxter equation}
\vspace{.15cm}
\centerline{September 2005}
\end{titlepage}

\renewcommand{\thefootnote}{\arabic{footnote}}

\section{Introduction}

The Yang-Baxter equation is undoubtedly the corner stone of the theory of two dimensional
integrable systems of statistical mechanics and quantum field theory. It is frequently viewed
as an operator relation for a matrix $R_{ab}(x)$  defined on the tensor product 
of two $N$-dimensional vectors spaces
$V_{a}$ and $V_{b}$, which reads
\begin{equation}
\label{yb}
R_{12}(x_1) R_{13}(x_1 x_2) R_{23}(x_2)=R_{23}(x_2) R_{13}(x_1 x_2) R_{12}(x_1),
\end{equation}
where $x_i=e^{\lambda_i}$ are arbitrary multiplicative spectral parameters.

The elements of the $R$-matrix $R_{ab}(x)$ can be thought of either as the Boltzmann weights of
vertex models in statistical mechanics \cite{BA} or as factorizable scattering amplitudes between
particles in relativistic field theories \cite{ZAM}. Therefore, the search for solutions of the
Yang-Baxter equation is indeed a central issue in the field of exactly solvable models. Unfortunately,
a complete classification of the solutions of the Yang-Baxter equation is so far beyond our reach.
An important class of solutions is denominated trigonometric $R$-matrices which contain
an extra free parameter $q$ besides the spectral parameter. An approach to derive such $R$-matrices has its roots
on the possibility of performing an appropriate $q$-deformation in a given classical Lie algebra $\cal{G}$ 
\cite{FA,DRI,JIQ}.
This method, the $U_q[\cal{G}]$ quantum group framework, permits us in principle to associate 
a fundamental trigonometric $R$-matrix to each Lie algebra \cite{JI,BAZ} or Lie superalgebra \cite{BAZ1}.
In particular, the $R$-matrices expressions in terms
of the standard Weyl matrices have been known since two decades ago for all non-exceptional Lie algebras
\cite{JI}.   
Similar statement can not be made for superalgebras since the most general results are
still concentrated on the $U_q[sl(n|m)^{(1)}]$ symmetry \cite{CHA}. Representative examples for other superalgebras
have been investigated for particular supergroup symmetries and for instance can be found in refs. \cite{SA,LIN}.
In fact, attempts to systematically carry on the above program for superalgebras \cite{GO,GO1} have
encountered serious technical obstacles to be  overcome   before explicit expressions could be
written down.

In spite of these difficulties, some progresses 
have recently been made towards to the presentation of explicit
expressions for the $R$-matrices based on general classes of superalgebras \cite{WM,LIN1}. 
We have for instance exhibited \cite{WM} 
the $R$-matrices associated to the $U_{q}[sl(r|2m)^{(2)}]$, $U_{q}[osp(r|2m)^{(1)}]$ and
$U_{q}[osp(r=2n|2m)^{(2)}]$ quantum superalgebras in terms of the Weyl matrices. This step makes possible 
the statistical mechanics interpretation of these systems and has offered suitable expressions to perform 
the corresponding transfer matrices diagonalization. These results, however,  have been obtained 
for a specific grading of the Grassmann parities  by means of brute force
analysis  of the respective $U_q[\cal{G}]$ intertwiner operators.

The purpose of this paper is to elaborate further on our previous results \cite{WM} by first 
unveiling the algebraic structure that is behind   
the $U_{q}[sl(r|2m)^{(2)}]$, $U_{q}[osp(r|2m)^{(1)}]$ and
$U_{q}[osp(r=2n|2m)^{(2)}]$ $R$-matrices. We will argue  that 
these two-dimensional solvable lattice models are intimately
connected with the representations of the so-called Birman-Wenzl-Murakami algebra \cite{BWM,RE}. 
This relationship allows us to generalize these $R$-matrices for a more general class of gradings
and with a considerable amount of free-parameters related to possible multiparametric extension of
quantum algebras \cite{RES}. We believe that this study 
throw new light on the classification problem of the
fundamental trigonometric vertex models 
having both bosonic and fermionic degrees of freedom. In fact,
it
makes possible  the  derivation of novel families of such  
solvable models whose existence have not even been 
predicted before by means of the quantum group 
framework \cite{BAZ1,GO,GO1}.

This paper has been organized as follows. In section 2 we introduce multiparametric
representations of the braid algebra motived on the  structure of the
$U_{q}[sl(r|2m)^{(2)}]$, $U_{q}[osp(r|2m)^{(1)}]$ and
$U_{q}[osp(r=2n|2m)^{(2)}]$ $R$-matrices previously obtained by us \cite{WM}. These generalized 
representations are shown to
satisfy 
the Birman-Wenzl-Murakami algebra for a variety of grading choices.
We reintroduce the spectral parameter via the baxterization procedure \cite{JO,BAT} and  the multiparametric
$R$-matrices associated to the 
$U_{q}[sl(r|2m)^{(2)}]$, $U_{q}[osp(r|2m)^{(1)}]$ and
$U_{q}[osp(r=2n|2m)^{(2)}]$ superalgebras are presented. This
permits us in section
3 to find new representations of the dilute version of the Birman-Wenzl-Murakami algebra \cite{GRI}.
The study of the corresponding baxterization leads us to 
two novel families of $R$-matrices such that each of them  produces
two distinct vertex models branches whose edge variables can be of bosonic
or fermionic types. 
These systems turn out to be highly non-trivial extensions of the
$U_q[D^{(2)}_{n+1}]$ vertex model \cite{JI,BAZ}. This is the case  even in the absence of fermionic edge
variables because the free-parameters produce by themselves a generalized structure for the Boltzmann  weights.
Our conclusions are summarized in section 4.
In Appendix A we describe a special form of the additional free parameters that is helpful
to make connections to Lie superalgebras.  In appendix B we present the crossing matrices of a
$R$-matrix exhibited in section 3. 

\section{The braid-monoid algebra}

The braid-monoid algebra \cite{RE} is generated by the identity $I$, the braid operator $b_i^+$ and
its inverse $b_i^{-}$ as well as  the monoid 
$E_i$. The index $i$ represents for instance the $i$-$\it th$ site of a one-dimensional lattice of length $L$. 
As usual the braid operators $b_i^{\pm}$ obey the
Artin braid group algebra \cite{AR},
\begin{eqnarray}
\label{ba}
&& b^{+}_{i} b^{-}_{i} = b^{-}_{i} b^{+}_{i} =I  \nonumber \\
&& b^{+}_{i} b^{+}_{j} = b^{+}_{j} b^{+}_{i} \;\;\;\;\;\;\;\; \mbox{for} \;\; |i-j| \geq 2  \nonumber \\
&& b^{+}_{i} b^{+}_{i + 1} b^{+}_{i} = b^{+}_{i + 1} b^{+}_{i} b^{+}_{i + 1}.
\end{eqnarray}

On the other hand the monoid $E_i$ is a  Temperley-Lieb operator \cite{TL} 
subjected to the relations,
\begin{eqnarray}
\label{tl}
&& E_{i} E_{j} = E_{j} E_{i} \;\;\;\;\;\;\;\; \mbox{for} \;\; |i-j| \geq 2  \nonumber \\
&& E_{i}^{2} = Q E_{i} \\
&& E_{i} E_{i \pm 1} E_{i} = E_{i}, \nonumber
\end{eqnarray}
where $Q$ is a complex parameter.

The braid group and the Temperley-Lieb algebra can be combined together into a single two parameters
algebra
provided the following additional relations are satisfied,
\begin{eqnarray}
&& b^{+}_{i} E_{i} = E_{i} b^{+}_{i} = \omega E_{i} \nonumber \\
&& b^{+}_{i} E_{j} = E_{j} b^{+}_{i} \;\;\;\;\;\;\;\; \mbox{for} \;\; |i-j| \geq 2  \\
&& b^{+}_{i \pm 1} b^{+}_{i} E_{i \pm 1} = E_{i} b^{+}_{i \pm 1} b^{+}_{i} = E_{i} E_{i+1}, \nonumber
\label{extra}
\end{eqnarray}
where $\omega$ is another complex parameter.

In what follows we shall argue that the  multiparametric  
$U_{q}[sl(r|2m)^{(2)}]$, $U_{q}[osp(r|2m)^{(1)}]$ and
$U_{q}[osp(r=2n|2m)^{(2)}]$ $R$-matrices can be derived from the representations
of a quotient 
of the braid-monoid algebra  denominated
Birman-Wenzl-Murakami algebra \cite{BWM}.   
The idea is first  to investigate suitable
asymptotic limits of the corresponding $R$-matrices given by us for 
a specific grading \cite{WM}.  This study reveals us that the main structure of the braid representations
can indeed be generalized to accommodate additional free-parameters and many distinct $Z_2$ grading possibilities. 

In order to do that lets us briefly recollect some basic definitions. We start by recalling that the 
vector spaces of these $R$-matrices are 
constituted 
of $r$ bosonic and $2m$ fermionic degrees of
freedom. A given $\alpha$-$\it th$ degree of freedom is distinguished by its Grassmann parity $p_{\alpha}$,
\begin{equation}
p_{\alpha}=\cases{
0 \;\;\; \mbox{for} \;\; \alpha \;\; \mbox{bosonic } \;\; \cr
1 \;\;\; \mbox{for} \;\; \alpha \;\; \mbox{fermionic }. \;\; \cr }
\end{equation}

In this situation 
the relationship between braid algebra (\ref{ba}) and the Yang-Baxter equation 
is made with the help of the following graded permutator,
\begin{equation}
P= \displaystyle \sum_{\alpha,\beta=1}^{N} (-1)^{p_{\alpha} p_{\beta}} e_{\alpha\beta}
\otimes e_{\beta\alpha},
\end{equation}
where $N=r+2m$ and $e_{\alpha\beta}$ denotes the  standard $N\times N$ Weyl matrices.

In fact, by defining a new matrix ${\check{R}}_{ab}(x)= P_{ab} R_{ab}(x)$ one can
rewrite Eq.(\ref{yb}) in a form that is not only insensitive to grading, namely
\begin{equation}
\label{ybr}
{\check{R}}_{12}(x_1) {\check{R}}_{23}(x_1 x_2) {\check{R}}_{12}(x_2)=
{\check{R}}_{23}(x_2) {\check{R}}_{12}(x_1 x_2) {\check{R}}_{23}(x_1) ,
\end{equation}
but also with a striking similarity with the braid algebra (\ref{ba}).

The braid representation can now be obtained from a given 
${\check{R}}_{ab}(x)$ by considering appropriate limits of the spectral parameter $x$ such that
Eq.(\ref{ybr}) becomes asymptotically independent of the variables $x_i$.  In  our case  this  can
be achieved by taking the following limits,
\begin{equation}
b^{\pm(l)} = \lim_{ \lambda \rightarrow \pm \infty} \left [ \theta_{\pm}(x=e^{\lambda})
\check{R}_{12}(x=e^{\lambda}) \right ]
\label{lim}
\end{equation}
where $\theta_{\pm}(x)$ are appropriate normalizations. The  upper index $l$  in 
$b^{\pm(l)}$  
anticipates the existence of two possible classes of braids to be described below.
Furthermore, 
the braid operator 
$b_{i}^{+(l)}$ and its inverse $b_{i}^{-(l)}$ follow directly from $b^{\pm(l)}$ by the standard 
construction,
\begin{equation}
\label{braidi}
b_{i}^{\pm (l)} = \bigotimes_{j=1}^{i-1} I_N~~
b^{\pm (l)}~~
\bigotimes_{j=i+2}^{L} I_N
\label{gene}
\end{equation}
where $I_N$ is the $N\times N$ identity matrix.  

It turns out that the most general  braid representations  
$b^{\pm(l)}$ we have found, that are compatible with the  $R$-matrices presented by us \cite{WM},
have the following  form,
\begin{eqnarray}
\label{mybraid}
b^{+(l)}= \sum_{\alpha \neq \alpha '}^{N} (-1)^{p_{\alpha}}q^{1-2p_{\alpha}}
e_{\alpha \alpha} \otimes e_{\alpha \alpha}
+\sum_{\stackrel{\alpha ,\beta=1}{\alpha \neq \beta ,\beta '}}^{N}
(-1)^{p_{\alpha} p_{\beta}} e_{\beta \alpha} \otimes e_{\alpha \beta} \nonumber \\
+ (q - \frac{1}{q}) \sum_{\stackrel{\alpha ,\beta =1}{\alpha <\beta, \; \alpha \neq \beta ' }}^{N}
e_{\alpha \alpha} \otimes e_{\beta \beta}
+\sum_{\alpha , \beta=1}^{N} a^{+(l)}_{\alpha \beta}
e_{\alpha ' \beta} \otimes e_{\alpha \beta '}
\end{eqnarray}
and
\begin{eqnarray}
\label{braidinv}
b^{-(l)}= \sum_{\alpha \neq \alpha '}^{N} (-1)^{p_{\alpha}}q^{-1+2p_{\alpha}}
e_{\alpha \alpha} \otimes e_{\alpha \alpha}
+\sum_{\stackrel{\alpha ,\beta=1}{\alpha \neq \beta ,\beta '}}^{N}
(-1)^{p_{\alpha} p_{\beta}} e_{\beta \alpha} \otimes e_{\alpha \beta} \nonumber \\
+(\frac{1}{q} - q ) \sum_{\stackrel{\alpha ,\beta =1}{\alpha <\beta, \; \alpha \neq \beta ' }}^{N}
e_{\beta \beta} \otimes e_{\alpha \alpha}
+\sum_{\alpha , \beta=1}^{N} {a}^{-(l)}_{\alpha \beta}
e_{\alpha ' \beta} \otimes e_{\alpha \beta '},
\end{eqnarray}
where $\alpha '=N+1-\alpha$. 

An interesting feature of the above proposal is that 
there exists some
freedom in fixing the coefficients 
${a}^{\pm(l)}_{\alpha \beta}$ for several choices of the Grassmann parities.  These grading possibilities are
those consonant  with the many possible $U(1)$ symmetries implicitly assumed 
in our construction (\ref{mybraid},\ref{braidinv})
of the braids. More specifically, the parities 
$p_{\alpha}$ are required to satisfy the following reflexion condition,
\begin{equation}
p_{\alpha}=p_{\alpha'} .
\label{pp}
\end{equation}

Taking condition (\ref{pp}) into account we found that
the coefficients $a_{\alpha \beta}^{\pm (l)}$ can be represented in terms of the following
general forms,
\begin{eqnarray}
\label{bcoef}
a^{+(l)}_{\alpha \beta}= \cases{
(\frac{1}{q} - q) \left[ \frac{\epsilon_{\alpha}^{(l)}}{\epsilon_{\beta}^{(l)}}
q^{t_{\beta}^{(l)}- t_{\alpha}^{(l)}} - \delta_{\alpha \; \beta '} \right] \;\;\;\;\;\; \alpha > \beta \cr
0 \;\;\;\;\;\;\;\;\;\;\;\;\;\;\;\;\;\;\;\;\;\;\;\;\;\;\;\;\;\;\;\;\;\;\;\;\;\;\;\;\;\;\;\;\;\;   \alpha < \beta \cr
1 \;\;\;\;\;\;\;\;\;\;\;\;\;\;\;\;\;\;\;\;\;\;\;\;\;\;\;\;\;\;\;\;\;\;\;\;\;\;\;   \alpha = \beta = \beta '  \cr
(-1)^{p_{\alpha}} q^{-1+2p_{\alpha}} \;\;\;\;\;\;\;\;\;\;\;\;\;\;\;\;\;\;\; \alpha=\beta \neq \beta '   \cr}, \nonumber \\
\\
a^{-(l)}_{\alpha \beta}= \cases{
(q - \frac{1}{q}) \left[ \frac{\epsilon_{\alpha}^{(l)}}{\epsilon_{\beta}^{(l)}}
q^{t_{\beta}^{(l)}- t_{\alpha}^{(l)}} - \delta_{\alpha \; \beta '} \right] \;\;\;\;\;\; \alpha < \beta \cr
0 \;\;\;\;\;\;\;\;\;\;\;\;\;\;\;\;\;\;\;\;\;\;\;\;\;\;\;\;\;\;\;\;\;\;\;\;\;\;\;\;\;\;\;\;\;\; \alpha > \beta \cr
1 \;\;\;\;\;\;\;\;\;\;\;\;\;\;\;\;\;\;\;\;\;\;\;\;\;\;\;\;\;\;\;\;\;\;\;\;\;\;\;  \alpha = \beta = \beta '  \cr
(-1)^{p_{\alpha}} q^{+1-2p_{\alpha}} \;\;\;\;\;\;\;\;\;\;\;\;\;\;\;\;\;\;\; \alpha=\beta \neq \beta ' \cr}. \nonumber
\end{eqnarray}

The remarked possibility of two different series of braids is therefore encoded in the variables
$\epsilon_{\alpha}^{(l)}$ and $t_{\alpha}^{(l)}$.   
The first family is defined for any integer value of $N$ and the respective parameters 
$\epsilon_{\alpha}^{(1)}$ and $t_{\alpha}^{(1)}$ satisfy the relations
\begin{eqnarray}
\label{le1}
\epsilon_{\alpha}^{(1)}=(-1)^{p_{\alpha}} \epsilon_{\alpha '}^{(1)} \;\;\;\;\;\;\; \mbox{and} \;\;\;\;\;
t_{\alpha}^{(1)} = t_{\alpha '}^{(1)} -2\left[p_{\alpha} + \frac{N}{2} - \alpha - 2\displaystyle{\sum_{\beta=\alpha}^{\left[\frac{N+1}{2}\right]} p_{\beta}} \right],
\end{eqnarray}
where $\alpha$ can take values on the interval $1 \leq \alpha < \left[  \frac{N+1}{2} \right]$.
We recall that $\left [ \frac{N+1}{2} \right ] $ denotes   the largest integer  less than $\frac{N+1}{2}$.

The second family of braid representations is valid only for $N$ even and the respective variables
$\epsilon_{\alpha}^{(2)}$ and $t_{\alpha}^{(2)}$ are then given by
\begin{eqnarray}
\label{le2}
\epsilon_{\alpha}^{(2)}=-(-1)^{p_{\alpha}} \epsilon_{\alpha '}^{(2)}
\;\;\;\;\;\;\; \mbox{and} \;\;\;\;\;
t_{\alpha}^{(2)} = t_{\alpha '}^{(2)} -2\left[p_{\alpha} + \frac{N}{2}+1 - \alpha - 2\displaystyle{\sum_{\beta=\alpha}^{\frac{N}{2}} p_{\beta}} \right],
\end{eqnarray}
where in this case $1 \leq \alpha \leq \frac{N}{2}$.

From expressions (\ref{le1},\ref{le2}) we conclude that each set of variables
$\epsilon_{\alpha}^{(l)}$ and $t_{\alpha}^{(l)}$  provides us the number of 
$\left[  \frac{N+1}{2} \right]$  free parameters. 
This
freedom is expected for braids related to representations of quantum algebras
because Hopf algebras  can accommodate  suitable multiparametric extensions \cite{RES}.
However, the explicit construction of universal $R$-matrices with such additional
free parameters is by no means a simple task, specially for superalgebras. 
Here we conjecture that the braids 
$b^{+(1)}$ and $b^{+(2)}$ are in direct correspondence 
with the multiparametric $U_q[osp(r|2m)^{(1)}]$ 
and $U_q[osp(r=2n|2m)^{(2)}]$ universal $R$-matrices, respectively. An evidence supporting this
conjecture is discussed in Appendix A.
Before proceeding we remark that such type of braids that mix both
bosonic and fermionic variables  have early been referred as ``non-standard'' braid
group representations \cite{COT,GE}. To our knowledge, however,
the expressions for the braids in terms of the standard Weyl matrices
(\ref{mybraid}) for general $N$ and with many arbitrary variables
$\epsilon_{\alpha}^{(l)}$ and $t_{\alpha}^{(l)}$ as well as their 
relationship with multiparametric
$R$-matrices invariant by superalgebras
are novel results in the literature.

We now turn our attention to the study of the eigenvalues structure of the
braids (\ref{mybraid}-\ref{le2}). In order to establish 
the connection with the representations of the Birman-Wenzl-Murakami  
algebra is important that such
braids have at most three distinct eigenvalues \cite{BAT}. By direct inspection
we conclude that $b^{+(l)}$ indeed satisfies the following cubic relation,
\begin{equation}
\label{eqc}
\left( b^{+(l)} + \frac{1}{q} I_N\otimes I_N \right) \left( b^{+(l)} - q I_N \otimes I_N \right)
\left( b^{+(l)} - \sigma_{l} I_N \otimes I_N \right) =0,
\end{equation}
where the third eigenvalue $\sigma_{l}$ is given by
\begin{equation}
\sigma_{l}=\cases{
q^{1-r+2m} \;\;\;\;\;\;\;\; \mbox{for} \; l=1  \cr
-q^{-1-2n+2m} \;\; \mbox{for} \; l=2 \cr }.
\end{equation}

The next step in order to close Birman-Wenzl-Murakami algebra is to assure
that the corresponding Temperley-Lieb operator $E^{(l)}$ is related to the braid $b^{\pm(l)}$
through the following identity,
\begin{equation}
\label{TLL}
E^{(l)} = I_N \otimes I_N  + \frac{b^{+(l)} - b^{-(l)}}{q^{-1}-q} \;\; .
\end{equation}

By substituting expressions
(\ref{mybraid},\ref{braidinv}) into Eq.(\ref{TLL}) we can therefore determine the explicit form
of the respective
monoid operator, namely
\begin{equation}
E^{(l)}= \sum_{\alpha , \beta}^{N} \frac{\epsilon_{\alpha}^{(l)}}{\epsilon_{\beta}^{(l)}}
q^{t_{\beta}^{(l)}- t_{\alpha}^{(l)}}
{e}_{\alpha ' \beta} \otimes {e}_{\alpha \beta '}   \;\; .
\label{mono}
\end{equation}

Our next remaining task is  to verify whether or not 
the braid $b^{+(l)}$ (\ref{mybraid}-\ref{le2}) 
and the respective monoid $E^{(l)}$ (\ref{mono}) 
satisfy the braid-monoid relations (\ref{ba},\ref{extra}). This
indeed occurs provided 
the complex parameters $Q$ and $\omega$ are set to assume the values
\begin{equation}
Q=1+\frac{\sigma_{l} - \sigma_{l}^{-1}}{q^{-1}-q}~~~\mathrm{and}~~~\omega=\sigma_{l}  \;\; .
\label{parqw}
\end{equation}

Lets us now discuss how to introduce the spectral parameter $x$ into these
braids representations so as to construct the corresponding solution $\check{R}_{ab}(x)$ of the
Yang-Baxter equation. The Birman-Wenzl-Murakami algebra is known 
to provide us the sufficient
conditions to perform the baxterization procedure \cite{BAT}.
It turns out that for each representation of this braid-monoid algebra one can find
two independent matrices $\check{R}^{(l,1)}(x)$ 
and $\check{R}^{(l,2)}(x)$ 
satisfying Yang-Baxter equation (\ref{ybr}). These solutions are
\begin{equation}
\label{baxbwm1}
\check{R}^{(l,k)} (x) = (\frac{1}{q} - q)x(x-\xi_{l}^{(k)}) I_N \otimes I_N
+ (x-1)(x-\xi_{l}^{(k)}) b^{+(l)} + (q -\frac{1}{q}) x (x-1) E^{(l)},
\end{equation}
where $\xi_{l}^{(k)}$ is given by
\begin{equation}
\label{qsi}
\xi_{l}^{(k)}=\cases{
-\frac{q}{\sigma_{l}} \;\;\;\;\;\;\;\; \mbox{for} \; k=1 \cr
\frac{1}{q \sigma_{l}} \;\;\;\;\;\;\;\;\;\; \mbox{for} \; k=2 \cr}.
\end{equation}

From the first sight one would think that these two types of baxterization, when $N$ even and
$N$ odd are considered separately, would in principle
lead us to six different solvable vertex models. This, however, is not
the case because the $R$-matrices $\check{R}^{(1,1)}(x)$ and $\check{R}^{(2,2)}(x)$
for $N$ even coincide, after gauge transformations are performed.
Therefore, we have altogether five different $R$-matrices that can be
obtained by direct substitution of Eqs.(\ref{mybraid},\ref{mono}) into Eq.(\ref{baxbwm1}).
After some cumbersome simplifications, their expressions in terms
of the standard Weyl matrices are given by
\begin{eqnarray}
\label{rmatrix}
&& {\check{R}}^{(l,k)}(x) = \nonumber \\
&& \sum_{\stackrel{\alpha=1}{\alpha \neq \alpha'}}^{N}
(x-\xi_{l}^{(k)})(x^{1-p_{\alpha}} - q^2 x^{p_{\alpha}}) {e}_{\alpha \alpha} \otimes {e}_{\alpha \alpha}
+q(x-1)(x-\xi_{l}^{(k)}) \sum_{\stackrel{\alpha ,\beta=1}{\alpha \neq \beta,\alpha \neq \beta'}}^{N}
(-1)^{p_{\alpha} p_{\beta}} {e}_{\beta \alpha} \otimes {e}_{\alpha \beta}  \nonumber \\
&& +(1-q^2)(x-\xi_{l}^{(k)}) \left[ x \sum_{\stackrel{\alpha ,\beta=1}{\alpha < \beta,\alpha \neq \beta'}}^{N} 
{e}_{\alpha \alpha} \otimes {e}_{\beta \beta}
+ \sum_{\stackrel{\alpha ,\beta=1}{\alpha > \beta,\alpha \neq \beta'}}^{N} 
{e}_{\alpha \alpha} \otimes {e}_{\beta \beta} \right]
 + \sum_{\alpha ,\beta =1}^{N} d_{\alpha, \beta}^{(l,k)} (x)
{e}_{\alpha' \beta} \otimes {e}_{\alpha \beta'}, \nonumber \\
\end{eqnarray}
such that  the functions $d_{\alpha,\beta}^{(l,k)}(x)$ are 
\begin{equation}
\label{dab}
d_{\alpha, \beta}^{(l,k)} (\lambda)=\cases{
\displaystyle  q(x -1)(x -\xi_{l}^{(k)}) + x(q^2 -1)(\xi_{l}^{(k)} -1) \;\;\;\;\;\;\;\;\;\;\;\;\;\;\;\;\;\;\; \alpha=\beta=\beta' \;\; \cr
\displaystyle  (x -1)\left[ (x -\xi_{l}^{(k)}) (-1)^{p_{\alpha}} q^{2 p_{\alpha}} +x(q^2 -1) \right] \;\;\;\;\;\;\;\; \;\;\;\;  \;\; \alpha=\beta \neq \beta' \;\; \cr
\displaystyle  (q^{2 }-1)\left[ \xi_{l}^{(k)}(x -1)\frac{\epsilon_{\alpha}^{(l)}}{\epsilon_{\beta}^{(l)}} 
q^{t_{\alpha}^{(l)}-t_{\beta}^{(l)}} -\delta_{\alpha ,\beta'} (x -\xi_{l}^{(k)}) \right] \;\;\;\;\;\;\; \; \;\; \alpha < \beta \;\;\cr
\displaystyle  (q^{2 }-1) x \left[ (x -1)\frac{\epsilon_{\alpha}^{(l)}}{\epsilon_{\beta}^{(l)}} 
q^{t_{\alpha}^{(l)}-t_{\beta}^{(l)}} -\delta_{\alpha ,\beta'} (x -\xi_{l}^{(k)}) \right] \;\;\;\;\; \;\;\;\;\;\; \;\; \alpha > \beta \;\;\cr } .
\end{equation}

At this point we stress that expressions (\ref{rmatrix},\ref{dab}) are valid in the general situation
when  the variables
$\epsilon_{\alpha}^{(l)}$ and $t_{\alpha}^{(l)}$ fulfill the relations (\ref{le1},\ref{le2}) and for the
variety of gradings satisfying condition (\ref{pp}). The possible relationship between such $R$-matrices
and the corresponding underlying quantum superalgebras is proposed in Table 1. 
This matching has been done by comparing Eqs.(\ref{rmatrix},\ref{dab}) with our previous $R$-matrices
results \cite{WM}. This comparison has also taken into account a symmetrical form for the variables
$\epsilon_{\alpha}^{(l)}$ and $t_{\alpha}^{(l)}$ described in Appendix A.
\begin{table}[h]
\begin{center}
\begin{tabular}{|c|c|c|} \hline
$\check{R}$-matrix  & Superalgebra & $\xi_{l}^{(k)}$ \\ \hline
$\check{R}^{(1,1)}(x)$  & $U_q[sl^{(2)}(r|2m)]$  & $-q^{r-2m}$ \\ \hline
$\check{R}^{(1,2)}(x)$  & $U_q[osp^{(1)}(r|2m)]$  & $q^{r-2m-2}$ \\ \hline
$\check{R}^{(2,1)}(x)$  & $U_q[osp^{(2)}(r=2n|2m)]$  & $q^{2n-2m+2}$ \\ \hline
\end{tabular}
\caption{The relationship between $\check{R}^{(l,k)}(x)$ and superalgebras. The expressions for the
parameters $\xi_l^{(k)}$ are also given.}
\end{center}
\end{table}

We believe  that these $R$-matrices together with Table 1 extend in a significative way our earlier results \cite{WM}
for solvable models based on superalgebras.  
In next section we shall see that we can profit even more from the approach described here.

\section{ The dilute braid-monoid algebra}

The dilute braid-monoid algebra \cite{GRI} turns out to be an interesting special case of the 
two colour generalization of the braid-monoid algebra \cite{PE}. The later algebra is generated
by coloured  braid $b_i^{\pm(a,b)}$ and monoid $E_i^{(a,b)}$ operators such that the label $a,b=1,2$ 
denotes the two possible colours.  The other elements are the projectors $P_i^{(a)}$ which
project onto the $a$-$\it th$ colour  at the $i$-$\it th$ 
site of a chain of size $L$. They satisfy the standard projectors
relations given by
\begin{equation}
\label{proj}
P_{i}^{(a)} P_{i}^{(b)} = \delta_{ab} P_{i}^{(a)}~~~ 
\sum_{a=1}^{2} P_{i}^{(a)} = I \;\; .
\end{equation}

In the dilute case one of the colours  plays the role of vacancy of a string in the usual braid-monoid
diagrams \cite{GRI}. This means that the corresponding representation of the subalgebra generated by the elements
related to this colour  is one-dimensional.  Choosing the second  colour $a=2$ as an empty string the non-trivial
braids entering in the dilute braid-monoid algebra are $b_i^{\pm(1,1)}$, $b_i^{+(1,2)}$ and $b_i^{+(2,1)}$. They
satisfy the following generalized braid group relations \footnote{Here it has been assumed that 
$b_{i}^{+(a,b)}=b_{i}^{-(a,b)}$ for  $a \neq b$ and 
$b_i^{\pm (2,2)}=E_i^{(2,2)}=p_i^{(2,2)} $.}
\begin{eqnarray}
\label{cbraid}
b_{i}^{-(a,b)} b_{i}^{+(b,a)} &=& b_{i}^{+(a,b)} b_{i}^{-(b,a)} = p_{i}^{(b,a)} \nonumber \\
b_{i}^{+(a,b)} b_{i+1}^{+(c,b)} b_{i}^{+(c,a)} &=& b_{i+1}^{+(c,a)} b_{i}^{+(c,b)} b_{i+1}^{+(a,b)},  
\end{eqnarray}
where the $p_i^{(a,b)}$ are composed projectors defined as $p_i^{(a,b)}=P_i^{(a)} P_{i+1}^{(b)}$.

By the same token the corresponding coloured monoid operators $E_i^{(1,1)}$, 
$E_i^{(1,2)}$ and $E_i^{(2,1)}$ are subjected to
extended Temperley-Lieb relations,
\begin{eqnarray}
\label{ctl}
E_{i}^{(a,b)} E_{i}^{(c,a)} &=& {Q^{(a)}} E_{i}^{(c,b)} \nonumber \\
E_{i}^{(a,b)} E_{i + 1}^{(a,a)} E_{i}^{(c,a)} &=& E_{i}^{(c,b)} p_{i +1}^{(c,a)} \nonumber \\
E_{i}^{(a,b)} E_{i - 1}^{(a,a)} E_{i}^{(c,a)} &=& E_{i}^{(c,b)} p_{i -1}^{(a,c)}.  
\end{eqnarray}

As usual it is assumed that any two of such generators acting at positions $i$ and $j$ with
$|i-j| \geq 2$ commute. Besides that, 
an additional set of relations among the braid and monoid generators are required to be satisfied, namely
\begin{eqnarray}
\label{ctwist}
b_{i}^{+(a,a)} E_{i}^{(b,a)} & =& \omega^{(a)} E_{i}^{(b,a)} \nonumber \\
E_{i}^{+(a,b)} b_{i}^{(a,a)} & =& \omega^{(a)} E_{i}^{(a,b)} \nonumber \\
b_{i + 1}^{+(c,b)} b_{i}^{+(c,b)} E_{i + 1}^{(a,b)} &=& E_{i}^{(a,b)} b_{i + 1}^{+(c,a)} b_{i}^{+(c,a)} = 
E_{i}^{(c,b)} E_{i +1}^{(a,c)} \nonumber  \\
b_{i - 1}^{+(b,c)} b_{i}^{+(b,c)} E_{i - 1}^{(a,b)} &=& E_{i}^{(a,b)} b_{i - 1}^{+(a,c)} b_{i}^{+(a,c)} = 
E_{i}^{(c,b)} E_{i-1}^{(a,c)}.
\end{eqnarray}

In analogy to section 2 the dilute Birman-Wenzl-Murakami algebra emerges as a quotient of the dilute
braid monoid algebra (\ref{proj}-\ref{ctwist}). As before this quotient demands further restrictions
between the braids $b_i^{\pm(1,1)}$ and the Temperley-Lieb 
operators $E_i^{(1,1)}$ such as the analog of the
cubic relation (\ref{eqc}) given by,
\begin{equation}
\label{neqc}
\left( b_i^{+(1,1)} + \frac{1}{q} p^{(1,1)} \right) \left( b_i^{+(1,1)} - q p^{(1,1)} \right)
\left( b_i^{+(1,1)} - \omega^{(1)} p^{(1,1)} \right) =0.
\end{equation}
as well as the following polynomial relation for the monoid $E_i^{(1,1)}$,
\begin{equation}
\label{ntlo}
E_i^{(1,1)} = p^{(1,1)} + \frac{b_i^{+(1,1)} - b_i^{-(1,1)}}{q^{-1}-q} \;\; .
\end{equation}

A relevant feature of such quotient of the dilute braid-monoid algebra is that the operators
related to the first colour  $b_i^{\pm(1,1)}$, $E_i^{(1,1)}$ and $p_i^{(1,1)}$ close a subalgebra
of Birman-Wenzl-Murakami type. This suggests therefore that the braid-monoid operators constructed
in section 2 can be used as the starting point to obtain representations of the dilute version of
the Birman-Wenzl-Murakami algebra. More precisely, these representations can be found from our previous
results by first adding one extra bosonic degree of freedom, corresponding to the second colour, to
the original local space of states. As a consequence of that 
the action of a given operator $\hat{O}_i$ at the $i$-$\it th $
site is now given by
\begin{equation}
\label{elo}
\hat{O}_{i} = \bigotimes_{j=1}^{i-1} I_{N+1}~~
\hat{O}~~ 
\bigotimes_{j=i+1}^{L} I_{N+1}   \;\; .
\end{equation}

The fact that the second  colour has been chosen  to be trivially represented leads us to the 
following general expressions for the projectors \cite{GRI},
\begin{equation}
\label{ppro}
P^{(1)} = \sum_{\alpha=1}^{N} \bar{e}_{\alpha \; \alpha}~~~~~ 
P^{(2)} = \bar{e}_{N+1 \;N+1}, 
\end{equation}
where $\bar{e}_{\alpha \beta}$ are $(N+1) \times (N+1)$ Weyl matrices.

The respective representations for the braids $b^{\pm(l|1,1)}$ and monoids $E^{(l|1,1)}$ can then
formally be taken from Eqs.(\ref{mybraid},\ref{braidinv},\ref{mono}), where once again 
the upper index $l$ takes account of two possible classes of representations. More specifically, these 
operators
are now $(N+1)^2 \times (N+1)^2$ matrices whose explicit expressions are, 
\begin{eqnarray}
\label{braiddil}
b^{+(l|1,1)}= \sum_{\alpha \neq \alpha '}^{N} (-1)^{p_{\alpha}}q^{1-2p_{\alpha}}
\bar{e}_{\alpha \alpha} \otimes \bar{e}_{\alpha \alpha}
+\sum_{\stackrel{\alpha ,\beta=1}{\alpha \neq \beta ,\beta '}}^{N}
(-1)^{p_{\alpha} p_{\beta}} \bar{e}_{\beta \alpha} \otimes \bar{e}_{\alpha \beta} \nonumber \\
+ (q - \frac{1}{q}) \sum_{\stackrel{\alpha ,\beta =1}{\alpha <\beta, \; \alpha \neq \beta ' }}^{N}
\bar{e}_{\alpha \alpha} \otimes \bar{e}_{\beta \beta}
+\sum_{\alpha , \beta=1}^{N} a^{+(l)}_{\alpha \beta}
\bar{e}_{\alpha ' \beta} \otimes \bar{e}_{\alpha \beta '},
\end{eqnarray}
\begin{eqnarray}
b^{-(l|1,1)}= \sum_{\alpha \neq \alpha '}^{N} (-1)^{p_{\alpha}}q^{-1+2p_{\alpha}}
\bar{e}_{\alpha \alpha} \otimes \bar{e}_{\alpha \alpha}
+\sum_{\stackrel{\alpha ,\beta=1}{\alpha \neq \beta ,\beta '}}^{N}
(-1)^{p_{\alpha} p_{\beta}} \bar{e}_{\beta \alpha} \otimes \bar{e}_{\alpha \beta} \nonumber \\
+(\frac{1}{q} - q ) \sum_{\stackrel{\alpha ,\beta =1}{\alpha <\beta, \; \alpha \neq \beta ' }}^{N}
\bar{e}_{\beta \beta} \otimes \bar{e}_{\alpha \alpha}
+\sum_{\alpha , \beta=1}^{N} a^{-(l)}_{\alpha \beta}
\bar{e}_{\alpha ' \beta} \otimes \bar{e}_{\alpha \beta '},
\end{eqnarray}
\begin{equation}
E^{(l|1,1)}= \sum_{\alpha , \beta}^{N} \frac{\epsilon_{\alpha}^{(l)}}{\epsilon_{\beta}^{(l)}}
q^{t_{\beta}^{(l)}- t_{\alpha}^{(l)}}
{\bar{e}}_{\alpha ' \beta} \otimes {\bar{e}}_{\alpha \beta '}.
\label{monod}
\end{equation}

At this point it should be emphasized that the operators (\ref{braiddil}-\ref{monod}) together with the projector
$p^{(1,1)}$ close the dilute Birman-Wenzl-Murakami subalgebra as long as the parameters $Q^{(1)}=Q$
$\omega^{(1)}=\omega$ 
and $Q^{(2)}=\omega^{(2)}=1$.
The expressions for the mixed braids 
$b^{+(l|1,2)}$
and 
$b^{+(l|2,1)}$ follows almost directly from the definition of the projectors, namely
\begin{equation}
\label{rem}
b^{+(l|1,2)} = \sum_{\alpha=1}^{N} \bar{e}_{N+1 \; \alpha} \otimes \bar{e}_{\alpha \; N+1}
\;\;\;\;\;\; b^{+(l|2,1)}= \sum_{\alpha=1}^{N} \bar{e}_{\alpha \; N+1} \otimes \bar{e}_{N+1 \; \alpha}.
\end{equation}

In order to obtain the mixed Temperley-Lieb operators some extra amount of work is however necessary.
It turns out that they are given by
\begin{eqnarray}
E^{(l|2,1)} &=& \sum_{\alpha=1}^{N} \frac{\epsilon_{\alpha}^{(l)}}{\epsilon_{N+1}^{(l)}} q^{t_{N+1}^{(l)}-t_{\alpha}^{(l)}}
\bar{e}_{\alpha ' \; N+1} \otimes \bar{e}_{\alpha \; N+1}, 
\end{eqnarray}
\begin{eqnarray}
E^{(l|1,2)} &=& \sum_{\alpha=1}^{N} \frac{\epsilon_{N+1}^{(l)}}{\epsilon_{\alpha}^{(l)}} q^{t_{\alpha}^{(l)} - t_{N+1}^{(l)}}
\bar{e}_{N+1 \; \alpha} \otimes \bar{e}_{N+1 \; \alpha '} , 
\label{rem1}
\end{eqnarray}
where 
${\epsilon_{N+1}^{(l)}}$ and $t_{N+1}^{(l)}$ are arbitrary additional parameters associated to the dilution. 

Next we turn to the problem of constructing spectral parameter dependent $R$-matrices from the
above realizations of the dilute Birman-Wenzl-Murakami algebra. As before, every 
representation of this algebra can be baxterized to yield a solution of the
Yang-Baxter equation \cite{GRI}. The corresponding expression for $\check{R}^{(l)}(x)$ is
\begin{eqnarray}
\label{fgrimm}
\check{R}^{(l)}(x) &=& (\frac{1}{q} - q) \eta_{l} p^{(1,1)} + (x- \frac{1}{x})(\frac{x}{\tau_{l}} b^{+(l|1,1)}
-  \frac{\tau_{l}}{x} b^{-(l|1,1)}) +  (\frac{1}{q} - q) ( \frac{\tau_{l}}{x} -  \frac{x}{\tau_{l}} ) (p^{(1,2)} +
p^{(2,1)}) \nonumber \\
&-& \kappa_{1} (x- \frac{1}{x}) ( \frac{\tau_{l}}{x} -  \frac{x}{\tau_{l}} ) ( b^{(l|1,2)} +b^{(l|2,1)} )
+ \kappa_{2} (\frac{1}{q} - q) (x- \frac{1}{x}) (E^{(l|1,2)} +E^{(l|2,1)} ) \nonumber \\
&+& \left[ \eta_{l} (\frac{1}{q} - q) - (x- \frac{1}{x}) ( \frac{\tau_{l}}{x} -  \frac{x}{\tau_{l}} ) \right] p^{(2,2)},
\end{eqnarray}
where $\tau_{l}^{2}= \sigma_{l}$, $\eta_{l}=\tau_{l} -\tau_{l}^{-1}$ and
$\kappa_{1,2}=\pm 1$ are arbitrary signs.

Direct substitution of the dilute representations (\ref{ppro}-\ref{rem1}) 
in Eq.(\ref{fgrimm})
leads us to expressions for the $R$-matrices whose expected underlying $U(1)$ symmetries are difficult
to be recognized at first sight. These  charge conservations can however be made more explicit
by means of suitable unitary transformations that preserve the Yang-Baxter equation, namely
\begin{equation}
\check{\mathcal{R}}^{(l)} (x)=\left( S \otimes S \right)^{-1}
\check{R}^{(l)} (x) \left( S \otimes S \right),
\end{equation}
where $S$ is an invertible $(N+1) \times (N+1) $ matrix.

For 
even $N=2n+2m$ we have found that the appropriate matrix $S_{even}$ is given by
\begin{equation}
\label{eventr}
S_{even}=\sum_{\alpha =1}^{n+m} \bar{e}_{\alpha \; \alpha} + \sum_{\alpha=n+m+1}^{2n+2m} \bar{e}_{\alpha \; \alpha+1}
+ \bar{e}_{2n+2m+1 \; n+m+1},
\end{equation}
and that the corresponding transformed $R$-matrices 
$\check{\mathcal{R}}^{(1)}(x)$ and $\check{\mathcal{R}}^{(2)}(x)$  are closely 
related to those of the $U_{q}[osp^{(1)}(2n+1|2m)]$ and $U_{q}[sl^{(2)}(2n+1|2m)]$
superalgebras given in section 2, respectively. In fact, they can be made equivalent
by spectral parameter dependent gauge 
transformations and therefore they do not produce new vertex
models.

The situation for odd $N=2n+1+2m$ is fortunately much more interesting. In this case the matrix $S_{odd}=S_{1}. S_{2}$
where $S_{1}$ and $S_{2}$ are given by,
\begin{eqnarray}
\label{oddtr}
S_{1} &=& \sum_{\alpha=1}^{n+m+1} \bar{e}_{\alpha \; \alpha} + 
\sum_{\alpha=n+m+2}^{2n+2m+1} \bar{e}_{\alpha \; \alpha+1} +\bar{e}_{ 2n+2m+2 \; n+m+2 } \nonumber \\
\\
S_{2} &=& \sum_{\alpha=1}^{n+m} \bar{e}_{\alpha \; \alpha} + \sum_{\alpha=n+m+3}^{2n+2m+2} \bar{e}_{\alpha \; \alpha} \nonumber \\
&+& \frac{1}{\sqrt{2}} \left( \bar{e}_{n+m+1 \; n+m+1} + \bar{e}_{n+m+1 \; n+m+2} + \bar{e}_{n+m+2 \; n+m+1} - \bar{e}_{n+m+2 \; n+m+2}
\right). \nonumber
\end{eqnarray}

The associated $R$-matrix
$\check{\mathcal{R}}^{(1)}(x)$ for $N$ odd is indeed novel as compared to that of the vertex models described
in section 2. The explicit form of such
$R$-matrix turns out to be
\begin{eqnarray}
\label{rsuperd}
&&\check{\mathcal{R}}^{(1)} (x) = \nonumber \\
&& \sum_{\stackrel{\alpha \neq \bar{n}+1}{\alpha \neq \bar{n}+2} } (x^2 -\zeta^{2}) \left[
x^{2(1-\bar{p}_{\alpha})} - q^2 x^{2 \bar{p}_{\alpha}} \right] \bar{e}_{\alpha \; \alpha} \otimes \bar{e}_{\alpha \; \alpha}
+ q (x^2 -1)(x^2 -\zeta^2) \sum_{\stackrel{\alpha \neq \beta, \beta ''}{\alpha, \beta \neq \bar{n}+1, \bar{n}+2} } (-1)^{\bar{p}_{\alpha} \bar{p}_{\beta}}\bar{e}_{\beta \; \alpha} \otimes \bar{e}_{\alpha \; \beta} \nonumber \\
&+& \frac{1}{2} q (x^2 -1)(x^2 -\zeta^2) \sum_{\stackrel{\alpha \neq \beta, \beta ''}{\beta = \bar{n}+1,\bar{n}+2}}
\left [ \right. (1+\kappa_{1}) \left(\bar{e}_{\beta \; \alpha} \otimes \bar{e}_{\alpha \; \beta} + \bar{e}_{\alpha \; \beta} \otimes \bar{e}_{\beta \; \alpha} \right)  \nonumber \\
&+& \left. (1-\kappa_{1}) \left(\bar{e}_{\beta \; \alpha} \otimes \bar{e}_{\alpha \; \beta ''} 
+ \bar{e}_{\alpha \; \beta} \otimes \bar{e}_{\beta '' \; \alpha} \right) \right ]
+\sum_{\alpha, \beta \neq \bar{n}+1, \bar{n}+2 } g_{\alpha \beta} (x)
\bar{e}_{\alpha '' \; \beta} \otimes \bar{e}_{\alpha  \; \beta ''}  \nonumber \\
&-& (q^2-1) (x^2 -\zeta^2) \left[ \sum_{\stackrel{\alpha < \beta , \alpha \neq  \beta ''}{\alpha, \beta \neq \bar{n}+1, \bar{n}+2}}
+ x^2 \sum_{\stackrel{\alpha > \beta , \alpha \neq  \beta ''}{\alpha, \beta \neq \bar{n}+1, \bar{n}+2}} \right]
\bar{e}_{\beta \; \beta} \otimes \bar{e}_{\alpha \; \alpha} \nonumber \\
&-& \frac{1}{2} (q^2-1)(x^2- \zeta^2) [ (x+1) \left( \sum_{\stackrel{\alpha < \bar{n} +1}{\beta = \bar{n}+1,\bar{n}+2}}
+ x  \sum_{\stackrel{\alpha > \bar{n} +2}{\beta = \bar{n}+1,\bar{n}+2}} \right)
\left( \bar{e}_{\beta \; \beta} \otimes \bar{e}_{\alpha \; \alpha}  +
\bar{e}_{\alpha '' \; \alpha ''} \otimes \bar{e}_{\beta '' \; \beta ''}  \right) \nonumber \\
&+& (x-1) \left( -\sum_{\stackrel{\alpha < \bar{n} +1}{\beta = \bar{n}+1,\bar{n}+2}}
+ x  \sum_{\stackrel{\alpha > \bar{n} +2}{\beta = \bar{n}+1,\bar{n}+2}} \right)
\left( \bar{e}_{\beta '' \; \beta} \otimes \bar{e}_{\alpha  \; \alpha}  +
\bar{e}_{\alpha '' \; \alpha ''} \otimes \bar{e}_{\beta '' \; \beta ''}  \right) ] \nonumber \\
&+& \frac{1}{2} \sum_{\stackrel{\alpha \neq \bar{n}+1, \bar{n}+2}{\beta=\bar{n}+1, \bar{n}+2}}
[ b_{\alpha}^{+}(x) \bar{e}_{\alpha '' \; \beta} \otimes \bar{e}_{\alpha  \; \beta ''}
+ \bar{b}_{\alpha}^{+} (x) \bar{e}_{\beta  \; \alpha ''} \otimes \bar{e}_{\beta '' \; \alpha}
+ b_{\alpha}^{-} (x) \bar{e}_{\alpha ''  \; \beta} \otimes \bar{e}_{\alpha  \; \beta} \nonumber \\
&+& \bar{b}_{\alpha}^{-}(x) \bar{e}_{\beta \; \alpha ''} \otimes \bar{e}_{\beta \; \alpha} ]
+\sum_{\alpha= \bar{n} +1,\bar{n}+2} [ c^{+} (x) \bar{e}_{\alpha '' \; \alpha} \otimes \bar{e}_{\alpha  \; \alpha ''}
+ c^{-} (x) \bar{e}_{\alpha \; \alpha} \otimes \bar{e}_{\alpha \; \alpha } \nonumber \\
&+&  d^{+} (x) \bar{e}_{\alpha '' \; \alpha ''} \otimes \bar{e}_{\alpha  \; \alpha }
+ d^{-} (x) \bar{e}_{\alpha \; \alpha ''} \otimes \bar{e}_{\alpha \; \alpha ''} ] \nonumber \\
&+&  \frac{1}{2} \kappa_{2} \zeta (q^2 -1)x(x^2 -1) {\cal{F}}_{-} \sum_{\alpha , \beta = \bar{n}+1, \bar{n}+2}
(-1)^{\alpha - \beta} \left( \bar{e}_{\beta \; \alpha } \otimes \bar{e}_{\alpha \; \beta ''}
+ \bar{e}_{\alpha \; \beta '' } \otimes \bar{e}_{\beta \; \alpha }
\right),
\end{eqnarray}
where $\alpha '' = N+2 - \alpha$, $ \bar{n}=n+m $ and $ \zeta=q^{n-m} $.

The  Boltzmann weights $g_{\alpha \; \beta}(x)$, $b_{\alpha}^{\pm}(x)$, $\bar{b}_{\alpha}^{\pm}(x)$, $c^{\pm}(x)$ and
$d^{\pm}(x)$ are given by,
\begin{eqnarray}
\label{bw}
g_{\alpha \; \beta}(x)  &=&\cases{
(x^2-1)\left[ (x^2 - \zeta^2) (-1)^{\bar{p}_{\alpha}} q^{2 \bar{p}_{\alpha}} + x^2 (q^2-1) \right] \;\;\;\;\;\;\;\;\; \alpha= \beta \cr
(q^2-1)\left[ \zeta^2 (x^2-1) \frac{\bar{\epsilon}_{\alpha}}{\bar{\epsilon}_{\beta}} q^{\bar{t}_{\alpha}-\bar{t}_{\beta}} -
\delta_{\alpha \; \beta ''} (x^2 - \zeta^2) \right] \;\;\;\;\;\; \alpha < \beta \cr
(q^2-1)x^2 \left[ (x^2-1) \frac{\bar{\epsilon}_{\alpha}}{\bar{\epsilon}_{\beta}} q^{\bar{t}_{\alpha}-\bar{t}_{\beta}} -
\delta_{\alpha \; \beta ''} (x^2 - \zeta^2) \right] \;\;\;\;\;\; \alpha > \beta \cr}  \\
\nonumber \\
b^{\pm}_{\alpha}(x) &=&\cases{
\mp \frac{\epsilon_{\alpha}^{(1)}}{\epsilon_{N+1}^{(1)}} q^{\tilde{t}_{\alpha}^{(1)}} (q^2-1)(x^2-1)(x \kappa_{2} \mp \frac{\epsilon_{N+1}^{(1)}}{\epsilon_{\bar{n}+1}^{(1)}} q^{t_{N+1}^{(1)} - t_{\bar{n} +1}^{(1)}} \zeta) \;\;\;\;\;\;\;\;\;\;\;\;\;\;\; \alpha < \bar{n}+1 \cr
(-1)^{p_{\alpha}} \frac{\epsilon_{\alpha ''}^{(1)}}{\epsilon_{\bar{n}+1}^{(1)}} q^{\tilde{t}_{\alpha}^{(1)}} (q^2-1)(x^2-1) x (x \mp \kappa_{2} \frac{\epsilon_{\bar{n}+1}^{(1)}}{\epsilon_{N+1}^{(1)}} q^{t_{\bar{n}+1}^{(1)} - t_{N+1}^{(1)}} \zeta) \;\;\;\;\;\; \alpha > \bar{n}+2 \cr} \\
\nonumber \\
\bar{b}^{\pm}_{\alpha}(x) &=&\cases{
\mp (-1)^{p_{\alpha}} \frac{\epsilon_{N+1}^{(1)}}{\epsilon_{\alpha}^{(1)}} q^{\tilde{t}_{\alpha}^{(2)}} (q^2-1)(x^2-1)(x \kappa_{2} \mp \frac{\epsilon_{\bar{n}+1}^{(1)}}{\epsilon_{N+1}^{(1)}} q^{t_{\bar{n}+1}^{(1)} - t_{N +1}^{(1)}} \zeta) \;\;\;\;\; \alpha < \bar{n}+1 \cr
\frac{\epsilon_{\bar{n}+1}^{(1)}}{\epsilon_{\alpha ''}^{(1)}} q^{\tilde{t}_{\alpha}^{(2)}} (q^2-1)(x^2-1) x (x \mp \kappa_{2} \frac{\epsilon_{N+1}^{(1)}}{\epsilon_{\bar{n}+1}^{(1)}} q^{t_{N+1}^{(1)} - t_{\bar{n}+1}^{(1)}} \zeta) \;\;\;\;\;\;\;\;\;\;\;\;\;\;\;\;\; \alpha > \bar{n}+2 \cr} \\
\end{eqnarray}
\begin{eqnarray}
\label{cco}
c^{\pm} (x) &=& \pm \frac{1}{2} (q^2-1)(\zeta+\kappa_{2} {\cal{F}}_{+})x(x \mp 1)\left[ x \kappa_{2}\frac{(\zeta {\cal{F}}_{+} + \kappa_{2})}{(\zeta+\kappa_{2} {\cal{F}}_{+})} \pm \zeta \right] + \frac{(1+\kappa_{1})}{2} q(x^2-1)(x^2 -\zeta^2) \nonumber \\
\end{eqnarray}
\begin{eqnarray}
\label{ddo}
d^{\pm} (x) &=& \pm \frac{1}{2} (q^2-1)(\zeta-\kappa_{2} {\cal{F}}_{+})x(x \pm 1)\left[ x \kappa_{2}\frac{(\zeta {\cal{F}}_{+} - \kappa_{2})}{(\zeta-\kappa_{2} {\cal{F}}_{+})} \pm \zeta \right] + \frac{(1-\kappa_{1})}{2} q(x^2-1)(x^2 -\zeta^2) \nonumber \\
\end{eqnarray}

The auxiliary variables ${\cal{F}}_{\pm}$, $\bar{\epsilon}_{\alpha}$, $\bar{t}_{\alpha}$, $\tilde{t}^{(1)}_{\alpha}$  and
$\tilde{t}^{(2)}_{\alpha}$ 
entering in the above weights definition depend directly on the additional parameters
$\epsilon^{(1)}_{\alpha}$ and $t^{(1)}_{\alpha}$ as follows,
\begin{equation}
{\cal{F}}_{\pm} = - \frac{1}{2} \left[ \frac{\epsilon_{\bar{n} +1}^{(1)}}{\epsilon_{N +1}^{(1)}} q^{t_{\bar{n}+1}^{(1)}-t_{N+1}^{(1)} } \pm \frac{\epsilon_{N +1}^{(1)}}{\epsilon_{\bar{n} +1}^{(1)}} q^{t_{N+1}^{(1)}-t_{\bar{n}+1}^{(1)} } \right]
\end{equation}
\begin{eqnarray}
\bar{\epsilon}_{\alpha} &=& \cases{
\displaystyle \epsilon_{\alpha}^{(1)} \;\;\; \;\;\;\; \;\; 1 \leq \alpha \leq  \bar{n}  \cr
\displaystyle \epsilon_{\alpha -1}^{(1)} \;\;\;\; \;\; \;\; \bar{n}+3 \leq \alpha \leq N+1  \cr }
\\ \nonumber \\
\bar{t}_{\alpha} &=& \cases{
\displaystyle t_{\alpha}^{(1)} \;\;\; \;\;\;\;\;\; 1 \leq \alpha \leq \bar{n}   \cr
\displaystyle t_{\alpha-1}^{(1)} \;\;\; \;\;\;\;\;  \bar{n}+3  \leq \alpha \leq N+1  \cr }
\end{eqnarray}
and
\begin{eqnarray}
\tilde{t}_{\alpha}^{(1)}&=&\cases{
\displaystyle t_{\alpha}^{(1)} - t_{N+1}^{(1)} + n - m \;\;\;\;\;\;\;\;\;\;\;\;\;\;\;\;\;\;\;\;\;\;\;\;\;\;\;\;\;\;\;\;\;\;\;\;\;\;\;\;\;\;\;\;\;\;\;\;\;\;\;\; 1 \leq \alpha \leq \bar{n} \cr
\displaystyle t_{\alpha ''}^{(1)} - t_{\bar{n}+1}^{(1)} + 2\alpha - 5 - 2\bar{n} - 2\bar{p}_{\alpha} - 4\sum_{\beta = \bar{n}+3}^{\alpha -1} \bar{p}_{\beta} \;\;\;\;\;\;\;\;\; \bar{n}+3 \leq \alpha \leq N+1  \cr}
\\ \nonumber \\
\tilde{t}_{\alpha}^{(2)}&=&\cases{
\displaystyle t_{N+1}^{(1)} - t_{\alpha}^{(1)} + 2\alpha - (n - m + 1) - 2\bar{p}_{\alpha} - 4\sum_{\beta=1}^{\alpha -1} \bar{p}_{\beta} \;\;\;\;\;\;\;\;\; 1 \leq \alpha \leq \bar{n} \cr
\displaystyle t_{\bar{n}+1}^{(1)} - t_{\alpha ''}^{(1)} \;\;\;\;\;\;\;\;\;\;\;\;\;\;\;\;\;\;\;\;\;\;\;\;\;\;\;\;\;\;\;\;\;\;\;\;\;\;\;\;\;\;\;\;\;\;\;\;\;\;\;\;\;\;\;\;\;\;\;\;\; \bar{n}+3 \leq \alpha \leq N+1  \cr}.
\end{eqnarray}

Finally, the renormalized parities $\bar{p}_{\alpha}$
are related to that of the section 2 by the following expression,
\begin{equation}
\label{newparity}
\bar{p}_{\alpha}=\cases{
p_{\alpha} \;\;\;\;\;\;\;\;\;\;\;\; 1 \leq \alpha \leq \bar{n} \cr
0 \;\;\;\;\;\;\;\;\;\;\;\;\;\; \alpha= \bar{n}+1 \cr
0 \;\;\;\;\;\;\;\;\;\;\;\;\;\; \alpha= \bar{n}+2 \cr
p_{\alpha-1} \;\;\;\;\;\;\;\;\; \bar{n}+3 \leq \alpha \leq N+1 \cr}.
\end{equation}

From 
Eqs. (\ref{rsuperd}-\ref{newparity}) one clearly notes that the above $R$-matrix has in fact two 
possible branches governed by the discrete parameter $\kappa_1=\pm 1$. This gives origin to two distinct
vertex models since the  structure of some of the Boltzmann  weights depend drastically on the sign of $\kappa_1$.
On the other hand, the parameter $k_2$ apparently
does not play such a relevant role in the $R$-matrix (\ref{rsuperd}). Indeed,
the transformation $\kappa_2 \rightarrow 
-\kappa_2$ followed  by similar reflexion in the variable $\zeta$ 
leaves the whole $R$-matrix (\ref{rsuperd}) invariant apart from a trivial sign change on the weights
$b_{\alpha}^{\pm}(x)$ and $\bar{b}_{\alpha}^{\pm}$.

To the best of our  knowledge the general multiparametric structure  
of the $R$-matrix (\ref{rsuperd}-\ref{newparity}) is new even when the fermionic degrees of freedom are 
absent. Though the basic form of  
$\check{\mathcal{R}}^{(1)} (x)$ for $\kappa_1=1$ with all $p_{\alpha}=0$ resembles that of the
$U_q[D^{(2)}_{n+1}]$ $R$-matrix given by Jimbo \cite{JI} there exists essential differences among
these $R$-matrices. A direct comparison reveals that our $R$-matrix presents extra relevant 
Boltzmann weights as compared to Jimbo's 
$U_q[D^{(2)}_{n+1}]$ vertex model \cite{JI} 
such as the last term of Eq.(\ref{rsuperd}). 
Besides that,  
the spectral parameter dependence
of some of the weights depends strongly on the additional variables $\epsilon_{\alpha}^{(1)}$ and 
$t_{\alpha}^{(1)}$. In fact, it is only for a fine tuning  between these extra parameters that 
all the above mentioned differences are canceled out. This appears to indicate that   
Jimbo's $U_q[D^{(2)}_{n+1}]$ $R$-matrix is a particular case and 
probably does not capture   
the most general structure admissible  in the
$U_q[D^{(2)}_{n+1}]$ quantum group deformations. Other indication of this fact occurs when 
one tries to solve the 
Jimbo's $U_q[D^{(2)}_{n+1}]$  vertex model by means of the quantum inverse scattering method \cite{PRE}.
One notices that the first nested Bethe ansatz for $n \geq 2$ is already governed by a multiparametric
$R$-matrix having   
more  Boltzmann weights entries than that of the 
Jimbo's $U_q[D^{(2)}_{n-1}]$ $R$-matrix.  Therefore, a consistent algebraic Bethe ansatz solution of these
systems will require  the
class of the multiparametric $R$-matrix 
exhibited here from the very beginning.

Yet another interesting property was found in the course of an explicit Yang-Baxter verification of Eqs.(\ref{rsuperd}-\ref{newparity}).
We observed that there exists a second integrable family differing from that defined by Eqs.(\ref{rsuperd}-\ref{newparity})
only in respect to the Boltzmann  weights 
$c^{\pm}(x)$ and $d^{\pm}(x)$. In other words, the whole structure  of the $R$-matrix (\ref{rsuperd}) as well as the form
of the weights $g_{\alpha ,\beta}(x)$, $b_{\alpha}^{\pm}(x)$ and $\bar{b}_{\alpha}^{\pm}(x)$
are kept unchanged except the
$c^{\pm}(x)$ and $d^{\pm}(x)$ weights.  For such second family  the spectral parameter dependence of 
the respective $c^{\pm}(x)$ and $d^{\pm}(x)$ weights  are
\begin{eqnarray}
\label{ccn}
c^{\pm} (x) &=& \pm \frac{1}{2} (q^2-1)(\zeta+\kappa_{2} {\cal{F}}_{+})x(x \mp 1)\left[ x \kappa_{2}\frac{(\zeta {\cal{F}}_{+} + \kappa_{2})}{(\zeta+\kappa_{2} {\cal{F}}_{+})} \pm \zeta \right] + \frac{(1-\kappa_{1})}{2} q(x^2-1)(x^2 -\zeta^2) \nonumber \\
\end{eqnarray}
\begin{eqnarray}
\label{ddn}
d^{\pm} (x) &=& \pm \frac{1}{2} (q^2-1)(\zeta-\kappa_{2} {\cal{F}}_{+})x(x \pm 1)\left[ x \kappa_{2}\frac{(\zeta {\cal{F}}_{+} - \kappa_{2})}{(\zeta-\kappa_{2} {\cal{F}}_{+})} \pm \zeta \right] + \frac{(1+\kappa_{1})}{2} q(x^2-1)(x^2 -\zeta^2) . \nonumber \\
\end{eqnarray}

Interesting enough we note that the weights  (\ref{ccn},\ref{ddn}) 
are related to the previous one (\ref{cco},\ref{ddo}) through the
reflexion $\kappa_1 \rightarrow -\kappa_1$. Here we stress that this transformation applies only 
for such specific weights subset. Therefore, one expects that Eq.(\ref{rsuperd}) with  weights
$c^{\pm}(x)$ and $d^{\pm}(x)$  given by either Eqs.(\ref{cco},\ref{ddo}) or 
Eqs.(\ref{ccn},\ref{ddn}) would provide us different $R$-matrices.
In fact, we have verified for some values of $L$ that 
the spectrum of the transfer matrices built from these two integrable families are
indeed unrelated. 

In order to emphasize the  extension of our results concerning the presence of fermionic degrees of freedom, 
it is 
convenient to present the $R$-matrix (\ref{rsuperd}-\ref{newparity}) for special choices of the 
additional parameters $\epsilon_{\alpha}^{(1)}$ and $t_{\alpha}^{(1)}$ such that 
Jimbo's $U_q[D^{(2)}_{n+1}]$ $R$-matrix is recovered when all $p_{\alpha}=0$. This occurs by choosing
the variables $\epsilon_{\alpha}^{(1)}$ and $t^{(1)}_{\alpha}$ for $1 \leq \alpha \leq N$ as described in Appendix A
as well as by setting
$\epsilon_{N+1}^{(1)}=-1$ and $t^{(1)}_{N+1}=n+m+1$. After carrying on the corresponding simplifications  in
Eqs.(\ref{rsuperd}-\ref{newparity}) 
we find that the $R$-matrix (\ref{rsuperd}) can be rewritten  as follows,
\begin{eqnarray}
\label{rsuperds}
&&{\bf \check{\mathcal{R}}}^{(1)} (x) = \nonumber \\
&& \sum_{\stackrel{\alpha \neq \bar{n}+1}{\alpha \neq \bar{n}+2} } (x^2 -\zeta^{2}) \left[
x^{2(1-\bar{p}_{\alpha})} - q^2 x^{2 \bar{p}_{\alpha}} \right] \bar{e}_{\alpha \; \alpha} \otimes \bar{e}_{\alpha \; \alpha}
+ q (x^2 -1)(x^2 -\zeta^2) \sum_{\stackrel{\alpha \neq \beta, \beta ''}{\alpha, \beta \neq \bar{n}+1,\bar{n}+2} }(-1)^{\bar{p}_{\alpha} \bar{p}_{\beta}} \bar{e}_{\beta \; \alpha} \otimes \bar{e}_{\alpha \; \beta} \nonumber \\
&+& \frac{1}{2} q (x^2 -1)(x^2 -\zeta^2) \sum_{\stackrel{\alpha \neq \beta, \beta ''}{\beta = \bar{n}+1,\bar{n}+2}}
[ (1+\kappa_{1}) \left(\bar{e}_{\beta \; \alpha} \otimes \bar{e}_{\alpha \; \beta} + \bar{e}_{\alpha \; \beta} \otimes \bar{e}_{\beta \; \alpha} \right) \nonumber \\
&+& (1-\kappa_{1}) \left(\bar{e}_{\beta \; \alpha} \otimes \bar{e}_{\alpha \; \beta ''} + \bar{e}_{\alpha \; \beta} \otimes \bar{e}_{\beta '' \; \alpha} \right) ]
+\sum_{\alpha, \beta \neq \bar{n}+1, \bar{n}+2 } \tilde{g}_{\alpha \beta} (x)
\bar{e}_{\alpha '' \; \beta} \otimes \bar{e}_{\alpha  \; \beta ''}  \nonumber \\
&-& (q^2-1) (x^2 -\zeta^2) \left[ \sum_{\stackrel{\alpha < \beta , \alpha \neq  \beta ''}{\alpha, \beta \neq \bar{n}+1, \bar{n}+2}}
+ x^2 \sum_{\stackrel{\alpha > \beta , \alpha \neq  \beta ''}{\alpha, \beta \neq \bar{n}+1, \bar{n}+2}} \right]
\bar{e}_{\beta \; \beta} \otimes \bar{e}_{\alpha \; \alpha} \nonumber \\
&-& \frac{1}{2} (q^2-1)(x^2- \zeta^2) [ (x+1) \left( \sum_{\stackrel{\alpha < \bar{n} +1}{\beta = \bar{n}+1,\bar{n}+2}}
+ x  \sum_{\stackrel{\alpha > \bar{n} +2}{\beta = \bar{n}+1,\bar{n}+2}} \right)
\left( \bar{e}_{\beta \; \beta} \otimes \bar{e}_{\alpha \; \alpha}  +
\bar{e}_{\alpha '' \; \alpha ''} \otimes \bar{e}_{\beta '' \; \beta ''}  \right) \nonumber \\
&+& (x-1) \left( -\sum_{\stackrel{\alpha < \bar{n} +1}{\beta = \bar{n}+1,\bar{n}+2}}
+ x  \sum_{\stackrel{\alpha > \bar{n} +2}{\beta = \bar{n}+1,\bar{n}+2}} \right)
\left( \bar{e}_{\beta '' \; \beta} \otimes \bar{e}_{\alpha  \; \alpha}  +
\bar{e}_{\alpha '' \; \alpha ''} \otimes \bar{e}_{\beta '' \; \beta ''}  \right) ] \nonumber \\
&+& \frac{1}{2} \sum_{\stackrel{\alpha \neq \bar{n}+1, \bar{n}+2}{\beta=\bar{n}+1, \bar{n}+2}}
\left[ \tilde{b}_{\alpha}^{+}(x) \left(\bar{e}_{\alpha '' \; \beta} \otimes \bar{e}_{\alpha  \; \beta ''}
+  \bar{e}_{\beta  \; \alpha ''} \otimes \bar{e}_{\beta '' \; \alpha} \right)
+ \tilde{b}_{\alpha}^{-} (x) \left( \bar{e}_{\alpha '' \; \beta} \otimes \bar{e}_{\alpha  \; \beta}
+ \bar{e}_{\beta \; \alpha ''} \otimes \bar{e}_{\beta \; \alpha} \right) \right] \nonumber \\
&+&\sum_{\alpha= \bar{n} +1,\bar{n}+2} \left[\tilde{c}^{+}_{\nu} (x) \bar{e}_{\alpha '' \; \alpha} \otimes \bar{e}_{\alpha \; \alpha ''}
+ \tilde{c}^{-}_{\nu} (x) \bar{e}_{\alpha \; \alpha} \otimes \bar{e}_{\alpha \; \alpha }
+ \tilde{d}^{+}_{\nu} (x) \bar{e}_{\alpha '' \; \alpha ''} \otimes \bar{e}_{\alpha \; \alpha }
+ \tilde{d}^{-}_{\nu} (x) \bar{e}_{\alpha \; \alpha ''} \otimes \bar{e}_{\alpha \; \alpha ''} \right] . \nonumber \\
\end{eqnarray}

The respective Boltzmann weights $\tilde{g}_{\alpha \; \beta}(x)$, $\tilde{b}^{\pm}_{\alpha}(x)$, 
$\tilde{c}_{\nu}^{\pm}(x)$ and
$\tilde{d}_{\nu}^{\pm}(x)$ are now given by
\begin{eqnarray}
\label{bws}
\tilde{g}_{\alpha \; \beta}(x)  &=&\cases{
(x^2-1)\left[ (x^2 - \zeta^2) (-1)^{\bar{p}_{\alpha}} q^{2 \bar{p}_{\alpha}} + x^2 (q^2-1) \right] \;\;\;\;\;\;\;\;\;\;\;\;\;\;\;\;\;\; \alpha= \beta \cr
(q^2-1)\left[ \zeta^2 (x^2-1) \frac{\breve{\epsilon}_{\alpha}}{\breve{\epsilon}_{\beta}} q^{\breve{t}_{\alpha}-\breve{t}_{\beta}} -
\delta_{\alpha \; \beta ''} (x^2 - \zeta^2) \right] \;\;\;\;\;\;\;\;\;\;\;\;\;\;\; \alpha < \beta \cr
(q^2-1)x^2 \left[ (x^2-1) \frac{\breve{\epsilon}_{\alpha}}{\breve{\epsilon}_{\beta}} q^{\breve{t}_{\alpha}-\breve{t}_{\beta}} -
\delta_{\alpha \; \beta ''} (x^2 - \zeta^2) \right] \;\;\;\;\;\;\;\;\;\;\;\;\;\;\; \alpha > \beta \cr} \\
\nonumber \\
\tilde{b}^{\pm}_{\alpha} (x) &=&\cases{
\pm \breve{\epsilon}_{\alpha} q^{\tilde{t}_{\alpha}} (q^2-1)(x^2-1)(x \kappa_{2} \pm \zeta) \;\;\;\;\;\;\;\;\;\;\;\;\;\;\;\;\;\;\;\;\;\;\;\;\;\;\;\;\;\; \alpha < \bar{n}+1 \cr
\breve{\epsilon}_{\alpha} q^{\tilde{t}_{\alpha}}(q^2-1)(x^2-1) x (x \kappa_{2} \pm \zeta) \;\;\;\;\;\;\;\;\;\;\;\;\;\;\;\;\;\;\;\;\;\;\;\;\;\;\;\;\;\;\; \alpha > \bar{n}+2 \cr} \\
\tilde{c}^{\pm}_{\nu} (x) &=& \pm \frac{1}{2} (q^2-1)(\zeta+\kappa_{2})x(x \mp 1)(x \kappa_{2} \pm \zeta) + \frac{(1+\nu \kappa_{1})}{2} q(x^2-1)(x^2 -\zeta^2) \\
\tilde{d}^{\pm}_{\nu} (x) &=& \pm \frac{1}{2} (q^2-1)(\zeta-\kappa_{2})x(x \pm 1)(x \kappa_{2} \pm \zeta) + \frac{(1- \nu \kappa_{1})}{2} q(x^2-1)(x^2 -\zeta^2)
\end{eqnarray}
where the lower index $\nu=\pm 1$ in the weights $c_{\nu}^{\pm}(x)$ and $d_{\nu}^{\pm}(x)$ indicates the
two possible families of models discussed previously. The explicit expressions for 
the variables $\breve{\epsilon}_{\alpha}$, $\breve{t}_{\alpha}$ and $\tilde{t}_{\alpha}$ are
\begin{eqnarray}
\breve{\epsilon}_{\alpha} &=& \cases{
\displaystyle (-1)^{-\frac{\bar{p}_{\alpha}}{2}} \;\;\;\;\;\;\;\;\;\;\;\;\; \;\; \;\; 1 \leq \alpha \leq  \bar{n}  \cr
\displaystyle 1 \;\;\;\;\;\; \;\;\;\;\;\; \;\;\;\;\;\;\;\;\;\;\;\;\;\;\;\; \alpha=\bar{n}+1  \cr
\displaystyle 1 \;\;\;\;\;\; \;\;\;\;\;\; \;\;\;\;\;\;\;\;\;\;\;\;\;\;\;\; \alpha=\bar{n}+2  \cr
\displaystyle (-1)^{\frac{\bar{p}_{\alpha}}{2}} \;\;\;\; \;\;\;\;\;\;\;\;\;\;\;\; \;\;\; \bar{n}+3 \leq \alpha \leq N+1  \cr }
\\ \nonumber \\
\breve{t}_{\alpha} &=& \cases{
\alpha + \left[ 1 - \bar{p}_{\alpha} +2\displaystyle{\sum_{\beta=\alpha }^{\bar{n}} \bar{p}_{\beta}} \right] \;\;\;\;\;\;\;\;\;\;\; \;\; 1 \leq \alpha \leq \bar{n}   \cr
\bar{n} +\frac{3}{2} \;\;\;\;\;\;\;\;\;\;\;\;\;\;\;\;\;\;\;\;\;\;\;\;\;\;\;\;\;\;\;\;\;\;\;\;\;\;\;\;\; \alpha= \bar{n}+1 \cr
\bar{n} +\frac{3}{2} \;\;\;\;\;\;\;\;\;\;\;\;\;\;\;\;\;\;\;\;\;\;\;\;\;\;\;\;\;\;\;\;\;\;\;\;\;\;\;\;\; \alpha= \bar{n}+2 \cr
\alpha - \left[ 1 -\bar{p}_{\alpha} +2\displaystyle{\sum_{ \beta= \bar{n}+3}^{\alpha} \bar{p}_{\beta}} \right] \;\;\;\;\;\;\;\; \;\;  \bar{n}+3  \leq \alpha \leq N+1  \cr }
\end{eqnarray}
\begin{eqnarray}
\label{tildet}
\tilde{t}_{\alpha}=\cases{
\displaystyle \alpha - \left[ \frac{1}{2} - \bar{p}_{\alpha} + 2 \sum_{\beta=1}^{\alpha} \bar{p}_{\beta} \right] \;\;\;\;\;\;\;\;\;\;\;\;\;\;\;\;\;\;\;\;\;\;\;\;\;\;\;\;\; 1 \leq \alpha \leq \bar{n} \cr
\displaystyle \alpha - \left[ \bar{n} + \frac{5}{2} - \bar{p}_{\alpha} + 2 \sum_{\beta=\bar{n}+3}^{\alpha} \bar{p}_{\beta}  \right]  \;\;\;\;\;\;\;\;\;\;\;\;\; \bar{n}+3 \leq \alpha \leq N+1  \cr}.
\end{eqnarray}

Now it is not difficult to recognize that expressions (\ref{rsuperds}-\ref{tildet})   
for the branch $\kappa_1=1$ and $\nu=1$  with
all $p_{\alpha}=0$ indeed reproduce the $U_q[D^{(2)}_{n+1}]$ $R$-matrix. This means that in general
the $R$-matrix (\ref{rsuperds}) should be considered as a non-trivial generalization 
of Jimbo's $U_q[D^{(2)}_{n+1}]$ vertex model when the respective edge variables admit both bosonic
and fermionic statistics. To our  knowledge such interesting possibility has not been predicted
before even in the realm of a powerful method such as the quantum supergroup formalism \cite{BAZ1,GO,GO1}.
To shed some light on the construction of the $R$-matrix (\ref{rsuperds}) in the context of quantum
superalgebras one can study its respective 
$q \rightarrow 1$ limit. By performing this analysis we found that the classical limit of the  
$R$-matrix (\ref{rsuperds}) with $\kappa_1=\kappa_2=\nu=1$ turns out to be the rational $osp(2n+2|2m)$
$R$-matrices \cite{KUM}.  Therefore it is plausible to suppose that the $R$-matrix (\ref{rsuperds}) could
be derived as a quantum deformation of the 
the $osp(2n+2|2m)$ Lie superalgebra with a given automorphism. It remains however the precise
identification of the order of the corresponding automorphism and this step has eluded us so far.

We would like to close this section by discussing useful properties satisfied by 
the $R$-matrix ${\bf{{\cal{R}}}}^{(1)}(x)=P {\bf{\check{\cal{{R}}}}}^{(1)}(x)$ where 
${\bf{\check{\cal{{R}}}}}^{(1)}(x)$  refers to the matrix given in Eq.(\ref{rsuperds}).
Besides regularity 
and unitarity this $R$-matrix satisfies the so-called $PT$ symmetry given by
\begin{equation}
P_{12}{\bf{\cal{R}}}^{(1)}_{12}(x)P_{12}=[{\bf{\cal{R}}}^{(1)}_{12}]^{st_1 st_2}(x) ,
\label{PT}
\end{equation}
where the symbol $st_k$ denotes the supertransposition in the space with index $k$. 
Yet another property is the
crossing symmetry, namely
\begin{equation}
\label{cross}
{\bf{\cal{R}}}^{(1)}_{12}(x)=\frac{\rho(x)}{\rho(\zeta/x)}
V_{1} [{\bf{\cal{R}}}^{(1)}_{12}]^{st_2}(\zeta/x)V^{-1}_{1},
\label{CRO}
\end{equation}
where $\rho(x)$ is a convenient normalization and $V$ is an
anti-diagonal matrix. The  explicit expressions for these quantities have been 
collected in Appendix B.

\section{Conclusions}

In this paper we have presented explicit representations of the Birman-Wenzl-Murakami algebra as well as
of its dilute generalization. The representations contain a considerable amount of
free parameters and the respective degrees of freedom can be of bosonic of fermionic type.
We argued that the corresponding braids should be related to the multiparametric universal 
$R$-matrices associated to the $U_q[osp(r|2m)^{(1)}]$ and $U_q[osp(r=2n|2m)^{(2)}]$ symmetries.

The baxterization of the representations of the Birman-Wenzl-Murakami algebra 
produced solutions of the Yang-Baxter equation invariant by the  
$U_q[osp(r|2m)^{(1)}]$,  $U_q[sl(r|2m)^{(2)}]$ 
and $U_q[osp(r=2n|2m)^{(2)}]$ quantum superalgebras. The dilute baxterization has leaded us to two
other families of $R$-matrices not previously foreseen by the framework of quantum supergroups. These systems
can in fact be considered as rather non-trivial  extensions of Jimbo's $U_q[D^{(2)}_{n+1}]$ $R$-matrix. This
occurs even when the fermionic variables are  absent
because the presence of extra parameters produces us the multiparametric
$U_q[D^{(2)}_{n+1}]$ $R$-matrix whose general structure was not known before. We noted that this 
knowledge is essential  to implement the algebraic Bethe ansatz for these vertex models in a consistent way.

Besides that, our study also pave the way to build the representations of the two-colour Birman-Wenzl-Murakami
algebra \cite{PE}. From any representation of this type one can in principle construct 
another $R$-matrices via the baxterization procedure. In view to what has been discussed above one expects
that new solvable vertex models could them be derived. It would be interesting to know the type of
lattice models with both bosonic and fermionic degrees of freedom that are obtained from this construction.
We hope to report on this problem in a future publication.

\section*{Acknowledgements}
The author W. Galleas thanks  FAPESP (Funda\c c\~ao de Amparo \`a Pesquisa do Estado de S\~ao Paulo)
for financial support. The work of M.J. Martins has been supported by the Brazilian Research Council-CNPq and FAPESP.

\newpage

\addcontentsline{toc}{section}{Appendix A}
\section*{\bf Appendix A: The symmetrical gauge}
\setcounter{equation}{0}
\renewcommand{\theequation}{A.\arabic{equation}}

In this appendix we briefly describe a symmetrical form for the variables
$\epsilon_{\alpha}^{(l)}$
and $t_{\alpha}^{(l)}$.  This choice of variables ensures us that the respective $R$-matrix becomes $PT$ invariant.
The idea is to explore the arbitrariness of Eqs.(\ref{le1},\ref{le2})
by fixing in a convenient way some of these variables. 

For the first family an appropriate choice of these variables on the interval
$1 \leq \alpha \leq  \frac{N+1}{2}$  will lead us to the following symmetrical structure,
\begin{eqnarray}
\epsilon_{\alpha}^{(1)} &=& \cases{
\displaystyle (-1)^{-\frac{p_{\alpha}}{2}} \;\;\;\;\;\;\;\;\;\;\;\;\;\;\;\;\;\; \;\; \;\; 1 \leq \alpha <  \frac{N+1}{2}  \cr
\displaystyle 1 \;\;\;\;\;\; \;\;\;\;\;\;\;\;\;\;\;\;\;\;\;\;\;\;\;\;\; \;\;\;\;\;\; \alpha=\frac{N+1}{2}  \cr
\displaystyle (-1)^{\frac{p_{\alpha}}{2}} \;\;\;\; \;\; \;\;\;\;\;\;\;\;\;\;\;\;\;\;\;\;\; \frac{N+1}{2}  < \alpha \leq N  \cr }
\\ \nonumber \\
t_{\alpha}^{(1)} &=& \cases{
\alpha + \left[ \frac{1}{2} -p_{\alpha} +2\displaystyle{\sum_{\alpha \leq \beta <  \frac{N+1}{2} } p_{\beta}} \right] \;\;\;\;\;\;\;\;\;\;\;\;\; \;\; 1 \leq \alpha < \frac{N+1}{2}   \cr
\frac{N+1}{2} \;\;\;\;\;\;\;\;\;\;\;\;\;\;\;\;\;\;\;\;\;\;\;\;\;\;\;\;\;\;\;\;\;\;\;\;\;\;\;\;\;\;\;\;\;\;\;\;\;\;\; \;\; \alpha =  \frac{N+1}{2}   \cr
\alpha - \left[ \frac{1}{2} -p_{\alpha} +2\displaystyle{\sum_{ \frac{N+1}{2}  < \beta \leq \alpha} p_{\beta}} \right] \;\;\;\;\;\;\;\;\;\;\;\;\; \;\;  \frac{N+1}{2}  < \alpha \leq N \cr } .
\end{eqnarray}

From our previous work \cite{WM} we see that this 
is exactly the form of the corresponding variables 
appearing  in  
the $U_{q}[osp(r|2m)^{(1)}]$ $R$-matrix.

Similarly, a suitable choice of the 
variables $\epsilon_{\alpha}^{(2)}$ and
$t_{\alpha}^{(2)}$ for 
$1 \leq \alpha \leq  \frac{N}{2}$ produces us the form
\begin{eqnarray}
\epsilon_{\alpha}^{(2)} &=& \cases{
\displaystyle (-1)^{-\frac{p_{\alpha}}{2}} \;\;\;\;\;\;\;\;\;\;\;\;\;\;\;\;\;\;\;\;\; 1 \leq \alpha \leq \frac{N}{2}  \cr
\displaystyle -(-1)^{\frac{p_{\alpha}}{2}} \;\;\;\;\;\;\;\;\;\;\;\;\;\;\;\;\;\;\;\; \frac{N}{2} +1 \leq \alpha \leq N \cr }
\\ \nonumber \\
t_{\alpha}^{(2)} &=& \cases{
\alpha - \left[ \frac{1}{2} +p_{\alpha} -2\displaystyle{\sum_{\beta=\alpha}^{\frac{N}{2}} p_{\beta}} \right] \;\;\;\;\;\;\;\;\; \;\;\;\;\;\;\;\;\;\;\;\; 1 \leq \alpha \leq \frac{N}{2}  \cr
\alpha + \left[ \frac{1}{2} +p_{\alpha} -2\displaystyle{\sum_{\beta=\frac{N}{2}+1}^{\alpha} p_{\beta}} \right] \;\;\;\;\; \;\;\;\;\;\;\;\;\;\;\;\; \frac{N}{2}+1 \leq \alpha \leq N \cr } \;\; ,
\end{eqnarray}
which is just that related with the 
$U_{q}[osp(r=2n|2m)^{(2)}]$
$R$-matrix given in ref.\cite{WM}.

The above results strongly suggest that $b^{+(1)}$ and 
$b^{+(2)}$  should be associated to the multiparametric 
$U_{q}[osp(r|2m)^{(1)}]$ and 
$U_{q}[osp(r=2n|2m)^{(2)}]$ 
universal $R$-matrices, respectively.

\addcontentsline{toc}{section}{Appendix B}
\section*{\bf Appendix B: Crossing symmetry }
\setcounter{equation}{0}
\renewcommand{\theequation}{B.\arabic{equation}}

The purpose here is to present
the explicit expressions for the
the normalization function $\rho(x)$ and the crossing matrix $V$.
The normalization is
\begin{equation}
\rho(x) = q(x^2-1)(x^2 - \zeta^2),
\end{equation}
while the only non-null entries of the matrix $V$ are the anti-diagonal elements $V_{\alpha \; \alpha ''}$, namely
\begin{equation}
V_{\alpha \; \alpha ''}=\cases{
\displaystyle (-1)^{\frac{\bar{p}_{\alpha}-1}{2}} \;\;\;\;\;\;\;\;\;\;\;\;\;\;\;\;\;\;\;\;\;\;\;\;\;\;\;\;\;\;\;\;\;\;\;\;\;\;\;\;\;\;\;\;\;\;\;\;\;\; \alpha =1 \cr
\displaystyle (-1)^{\frac{\bar{p}_{\alpha}-1}{2}} q^{\left[ \alpha - 1 - \bar{p}_{1} - \bar{p}_{\alpha} - 2\sum_{\beta=2}^{\alpha-1} \bar{p}_{\beta}  \right]} \;\;\;\;\;\;\;\;\;\;\;\;\;\;\;\; 1 < \alpha < \bar{n} +1 \cr
\displaystyle (-1)^{\frac{\bar{p}_{\alpha}-1}{2}} q^{\left[ \bar{n} - \frac{1}{2} - \bar{p}_{1} - \bar{p}_{\alpha} - 2\sum_{\beta=2}^{\bar{n}} \bar{p}_{\beta}  \right]} \;\;\;\;\;\;\;\;\;\;\;\;\;\;\;\; \alpha = \bar{n} +1 \cr
\displaystyle (-1)^{\frac{\bar{p}_{\alpha}-1}{2}} q^{\left[ \bar{n} - \frac{1}{2} - \bar{p}_{1} - \bar{p}_{\alpha} - 2\sum_{\beta=2}^{\bar{n}} \bar{p}_{\beta}  \right]} \;\;\;\;\;\;\;\;\;\;\;\;\;\;\;\; \alpha = \bar{n} +2 \cr
\displaystyle (-1)^{\frac{\bar{p}_{\alpha}-1}{2}} q^{\left[ \alpha - 3 - \bar{p}_{1} - \bar{p}_{\alpha} - 2\sum_{\beta=2}^{\alpha-1} \bar{p}_{\beta}  \right]} \;\;\;\;\;\;\;\;\;\;\;\;\;\;\;\; \bar{n}+2 < \alpha \leq N+1   \cr}
\end{equation}

{}

\end{document}